\documentclass[conference]{IEEEtran}
\ifCLASSINFOpdf

\else

\fi

\usepackage[table,usenames,dvipsnames]{xcolor}
\usepackage[T1]{fontenc}
\usepackage[tableposition=top]{caption}
\usepackage{rotating}
\usepackage{color, colortbl}
\usepackage{booktabs, array, dcolumn}
\usepackage{tikz}
\usepackage{graphicx}
\usepackage{pgf-umlsd}
\usepackage{multirow}
\usepackage{epsfig}
\usepackage{subfig}
\usepackage{pgfplotstable}
\usepackage{amsfonts}
\usepackage{relsize}
\usepackage{varwidth}
\usepackage{url}
\usepackage{amsmath}
\usepackage{amssymb}
\usepackage[normalem]{ulem}
\usepackage{diagbox}
\usepackage[english]{babel}
\usepackage{flushend}

\begin{document}
%
\title{A Delay-Tolerant Payment Scheme Based on the Ethereum Blockchain\vspace{-8mm}}



%
\author{\IEEEauthorblockN{Yining Hu,\IEEEauthorrefmark{1}\IEEEauthorrefmark{2} Ahsan Manzoor,\IEEEauthorrefmark{3}
Parinya Ekparinya,\IEEEauthorrefmark{4}
Madhusanka Liyanage,\IEEEauthorrefmark{3} \\ Kanchana Thilakarathna,\IEEEauthorrefmark{4} Guillaume Jourjon,\IEEEauthorrefmark{2} Aruna Seneviratne,\IEEEauthorrefmark{1}\IEEEauthorrefmark{2} and Mika E Ylianttila\IEEEauthorrefmark{3}}


\IEEEauthorblockA{\IEEEauthorrefmark{1}University of New South Wales, Australia, \IEEEauthorrefmark{2}Data61-CSIRO, Australia, \\ \IEEEauthorrefmark{3}University of Oulu, Finland, \IEEEauthorrefmark{4}University of Sydney, Australia}
	
\IEEEauthorblockA{Email: \IEEEauthorrefmark{1}firstname.lastname@data61.csiro.au,  \IEEEauthorrefmark{3}firstname.lastname@oulu.fi, \\ \IEEEauthorrefmark{4}pekp6601@uni.sydney.edu.au,  \IEEEauthorrefmark{4}kanchana.thilakarathna@sydney.edu.au}}



\maketitle

\begin{abstract}


Banking as an essential service can be hard to access in remote, rural regions where the network connectivity is intermittent. Although micro-banking has been made possible by SMS or USSD messages in some places, their security flaws and session-based nature prevent them from a wider adoption. Global level cryptocurrencies enable low-cost, secure and pervasive money transferring among distributed peers, but are still limited in their ability to reach more people in remote communities. 

We proposed to take advantage of the delay-tolerant nature of blockchains to deliver banking services to remote communities that only connect to the broader Internet intermittently. Using a base station that offers connectivity within the local area, regular transaction processing is solely handled by blockchain miners. The bank only joins to process currency exchange requests, reward miners and track user balances when the connection is available. By distributing the verification and storage tasks among peers, our system design saves on the overall deployment and operational costs without sacrificing the reliability and trustworthiness. Through theoretical and empirical analysis, we provided insights to system design, tested its robustness against network disturbances, and demonstrated the feasibility of implementation on off-the-shelf computers and mobile devices.


\end{abstract}


%


\section{Introduction}
Business operations in rural areas often face challenges associated with access, due to the lack of infrastructures. Communication infrastructure is one of the most crucial challenges, as an increasing number of services today rely on continuous network connectivity. 
However, connectivity in many of those difficult-to-access areas, even when available, is often intermittent. As a consequence, most of the pervasive services that are taken for granted cannot operate in these connectivity-restricted environments. This has led to the development of techniques, such as ad-hoc networks, and applications that can utilize the intermittent connection to provide some level of service~\cite{shah2003,Pentland2004} in the past. More recently, these regions are beginning to be served by community-run small-cell base stations for local connectivity, for example, Nokia Kuha~\cite{kuha} base stations that enable connectivity to the broader internet intermittently, but continuous interconnectivity for devices located within its coverage. 



One sector that has adopted different technologies available in these connectivity-restricted environments to deliver valuable services, is banking. 
For example, primitive banking services have been made possible via Short Message Service (SMS) or Unstructured Supplementary Service Data (USSD) of cellular networks. This is exemplified by the BAAC in Thailand~\cite{fitchett1999bank} and M-Pesa in Kenya and Tanzania~\cite{mas2010mobile}. The emergence of having interconnectivity within a given area, e.g., a village, with intermittent access to the main operator network, as well as decentralised services such as cryptocurrencies that do not require centralized control \cite{macdonald2016blockchains,mackenzie2015fintech}, has opened up possibilities for delivering improved banking services to connectivity-restricted areas.

Indeed, cryptocurrencies have received a significant attention in recent years and are regarded by many as a potential revolution of the banking industry \cite{macdonald2016blockchains,mackenzie2015fintech}. The technology behind, the blockchain, is secured by hard-coded software programs and enables peer democracy for transaction settlement. 
However, like most other pervasive services of today,  cryptocurrencies 
require continuous network connectivity to constantly exchange large volumes of data. For instance, a typical Bitcoin client uses around 200 MB of data per hour \cite{bandwidth_bitcoin}, while the Ethereum blockchain size has even surpassed the Bitcoin's in June 2017.\footnote{\url{http://www.altcointoday.com/ethereums-blockchain-size-surpasses-bitcoins-by-40/}} Therefore, to take advantage of the village-wide interconnectivity and the benefits of decentralisation, it is necessary to leverage the delay-tolerant characteristic of the blockchain and extend it to operate in intermittently connected environments.

This paper shows how this can be achieved by considering a scenario where  a community operates small-cell base stations for local connectivity with intermittent connection to the broader Internet, and a financial institution, i.e., a bank, that supplies, pays for and monitors equipment used in providing the service and
makes the following contributions:
\begin{enumerate}
	\item The design of a low-cost, accessible, reliable and secure payment scheme based on the Ethereum blockchain. 
	\item Mathematical models of the proposed system and an evaluation of the system performance under several operating conditions that confirms its feasibility. 
	\item Demonstrate the practical viability of proposed system through a prototype implementation using off-the-shelf laptops and mobile devices. 
	\item The first ever use of smart contracts for admission control, management of user accounts, mining rewards, and token creation. 
\end{enumerate}


The remainder of the paper is organized as follows: Section \ref{background} introduces technical details of blockchain. Section \ref{sec:sys_architecture} describes the system architecture and Section \ref{sec:modelling} provides models for the system design. Section \ref{sec:evaluation} presents the validation of our models and evaluation over local blockchain system. Section \ref{sec:implementation} demonstrates a prototype implementation with multiple test results. Section \ref{sec:related} discusses the related work, Section \ref{sec:discussion} highlights insights on future directions, and Section \ref{sec:conclusion} finally concludes the paper.


\section{Blockchain basics}
\label{background}
We base our system design on the Ethereum platform \cite{wood2014ethereum} that incorporates a Turing-complete scripting language for user-defined programs with a short block-generation time. We here explain key terminology and describe the Ethereum mechanisms. 

\subsection{Nodes and Their Connectivity}
A blockchain consists of a peer-to-peer (P2P) communication overlay network. In the case of Ethereum, each network node continuously attempts to connect to other nodes until they have peers. By default, Ethereum nodes use a gossip protocol to find out about other nodes. 

Once the overlay network is created, nodes on a blockchain may act as full nodes, miners or light nodes. All nodes contribute to the network connectivity and information (i.e., transactions and blocks) propagation. Miners and full nodes verify all transactions based on the signatures on them and the state changes they result in, and each keeps a complete transaction record. Light nodes operate in the Simplified Payment Verification (SPV) mode to only download block headers and verify and store transactions that are related to them. In addition, miners settle transactions through a process called mining as explained in Part \ref{subsec:mining}.

\subsection{Transactions, Blocks and Smart Contracts} 
\label{subsec:tx-blocks-sc}
The state of a blockchain is represented by peer interactions or transactions. For simplicity and efficiency, transactions are grouped together to be settled and immutably recorded in blocks. A user needs an externally controlled account (EOA) to send and receive transactions. An EOA has an ether balance and is controlled by the user's private key. Ethereum also uses smart contracts to form agreements among different entities. Smart contracts are identified by contract accounts, which are similar to EOAs but are associated with code that can be triggered by transactions or messages (calls) received from other contracts. Transactions can be sent between two EOAs, two contract accounts, or an EOA and a contract account. In the public Ethereum network, miners also collect transaction fees to make profit \cite{wood2014ethereum}. Transaction priority, i.e. how likely a transaction is to be picked up by miners, depends mostly on its value, age and the transaction fee attached. State of a blockchain is computed after every block \cite{bonneau2015sok}, and smart contracts are executed by all nodes when synchronization happens (cf. Part \ref{subsec:mining}). 

\subsection{The Blockchain}
\label{subsec:mining}
Nakamoto in his original paper proposed a proof-of-work (PoW) scheme to confirm transactions and select representation in majority decision making~\cite{nakamoto2008bitcoin}. PoW is also adopted by the current version of Ethereum as described below. 

\subsubsection{PoW Mining and Difficulty Control}
PoW mining is essentially the process of repeatedly calculating a hash value with an incrementing nonce until the hash is smaller than a target. Hashrate is the speed at which a miner computes hashes. Miners compete against each other in solving PoW puzzles and broadcast blocks to the rest of the network upon block creation for validation and synchronization. In addition to mining, miners also listen for new transactions and blocks discovered by others to prepare the next block. 

Difficulty is a measure of how difficult it is to solve a PoW puzzle. Difficulty adjustment over PoW puzzles leads to a stable block creation rate. In Ethereum the calculation is based on the block number, timestamp of the current block, and timestamp \& difficulty of its parent block. For Ethereum Homestead,\footnote{\url{http://ethdocs.org/en/latest/introduction/the-homestead-release.html}} the adjustment is always a multiple of \texttt{parent\textunderscore diff // 2048}, with parameter $a \in \left[-99, 1\right]$. We denote \texttt{block\textunderscore timestamp - parent\textunderscore timestamp} as $\delta t$, and summarize the adjustments in Table \ref{tb:diff_adjustment}. Difficulty adjustment is insignificant under a high network hashrate. 
\vspace{-4mm}
\begin{table}[!htbp]
	\centering
	\caption{PoW difficulty adjustments.} 
	\begin{tabular}{>{\centering\arraybackslash}p{3cm} | >{\centering\arraybackslash}p{4cm}} \specialrule{.12em}{1em}{0em}
		{\bf Timestamp difference} &  {\bf Adjustment} \\ \hline \hline
		$\delta t<10s$ & $a = 1$ \\ \hline
		$10s<= \delta t < 20s$ & Difficulty unchanged, $a = 0$ \\ \hline
		$20s<= \delta t < 1000s$ & Adjust downwards proportional to $\delta t$, $a \in [-99, -1]$ \\ \hline
		$\delta t >= 1000s$ & $a = -99$ \\ \hline
		\specialrule{.12em}{0em}{0em}
	\end{tabular}
	\label{tb:diff_adjustment}
\end{table}


\subsubsection{Consensus and Chain Growth}
Transmission delays may lead to stale blocks, or forks \cite{decker2013information} as multiple blocks could be created at the same time, and received by different nodes across the network. A block is attached to the chain once created, with its own block header pointing to the previous block header, and all the way back to the genesis block. All peers work out these inconsistencies by selecting the longest chain. Ethereum determines the longest chain based on the total difficulty of all the blocks.\footnote{Ethereum does not implement the full Greedy Heaviest-Observed Sub-Tree (GHOST) protocol proposed by Sompolinsky and Zohar \cite{sompolinsky2013accelerating}, instead, it uses a longest-chain rule with rewards for stale blocks \cite{gervais2016security}.} 

\subsubsection{Node Synchronization}
Network delays and node churns are common in blockchain systems. If one node restarts after going off-chain, it enquires its peers to obtain the latest version of the ledger. States stored in a particular block can only be accessed if the node synchronization has reached it.

\subsection{Coin Supply}
To compensate the resources consumed in mining, a fixed amount of new tokens, or mining reward, is generated upon the creation of blocks and paid to the winners. This rewarding scheme is necessary as it makes the blockchain more secure \cite{carlsten2016instability}, but the continuous creation of coins may result in inflation. 



\section{System Architecture}
\label{sec:sys_architecture}
The primary objective of the proposed system is to enable cash-less payment in a remote community that has constant connectivity within the local area but only intermittent connection to the broader Internet. Namely, the community connects to a bank's central network intermittently.
A network operator deploys a base station, similar to Nokia Kuha \cite{kuha} that forms a 4G LTE cellular network in the local region. The backhaul network could be a low-bandwidth satellite or a long-range microwave connection that does not provide robust service guarantees. 
Transaction processing is enabled using a private Ethereum blockchain \cite{wood2014ethereum} deployed by the bank. We assume some villagers would volunteer to participate in mining and be chosen by the bank. We do not use the default ether generation scheme to avoid inflation in the local economy, but instead, create our own digital currency, or Token, \footnote{\url{https://www.ethereum.org/token}} via a smart contract, for use within the community. The use of Tokens automatically imposes restrictions on users entering the system and makes it easier for the bank to manage user accounts.

\begin{figure}[!ht]
	\centering
	\subfloat[System architecture.]{\label{fig:sys-architecture}\includegraphics[width=1.1\columnwidth]{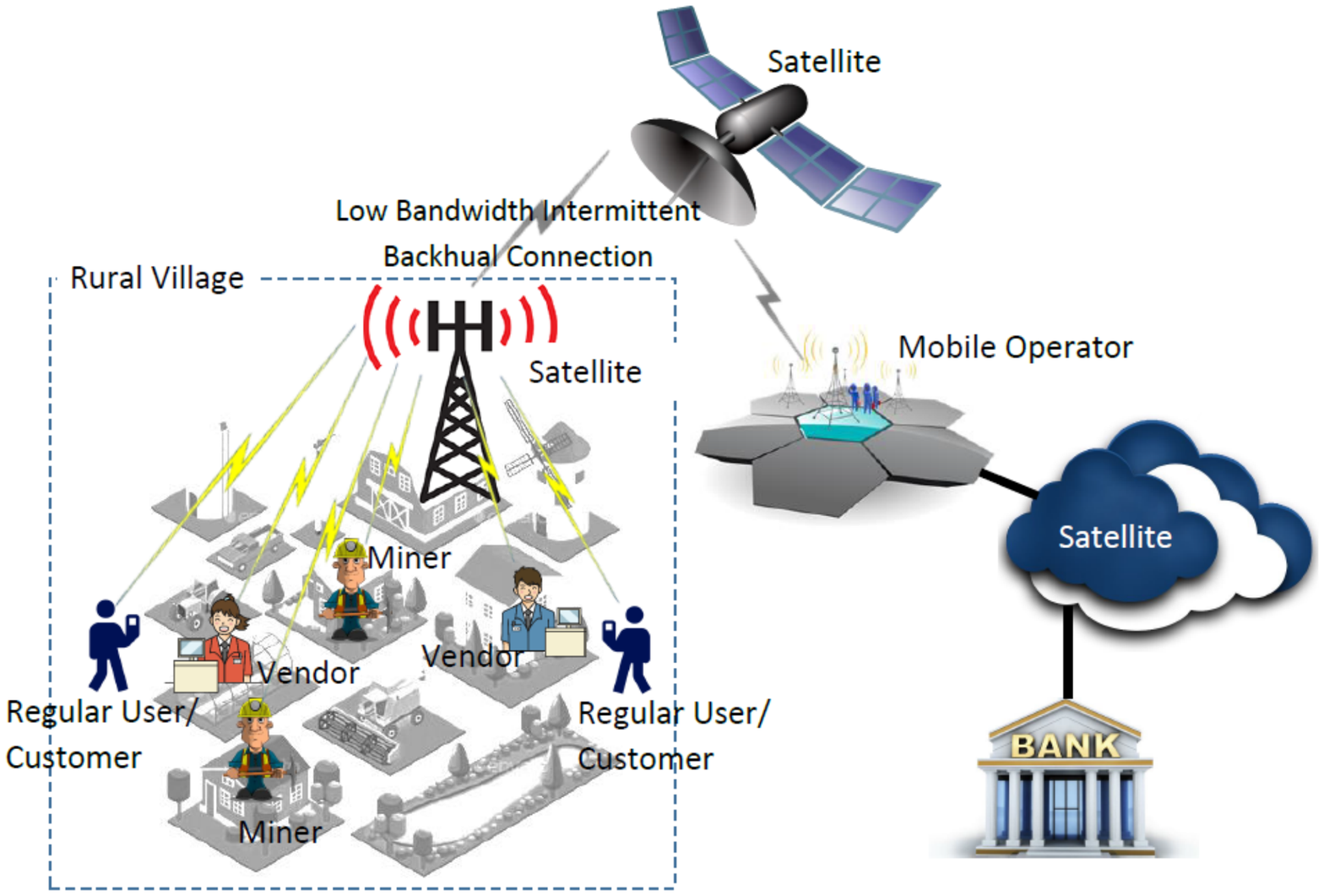}} 
	

	\subfloat[Transaction Flows.]{\label{fig:tx_flow}\includegraphics[width = 0.9\columnwidth]{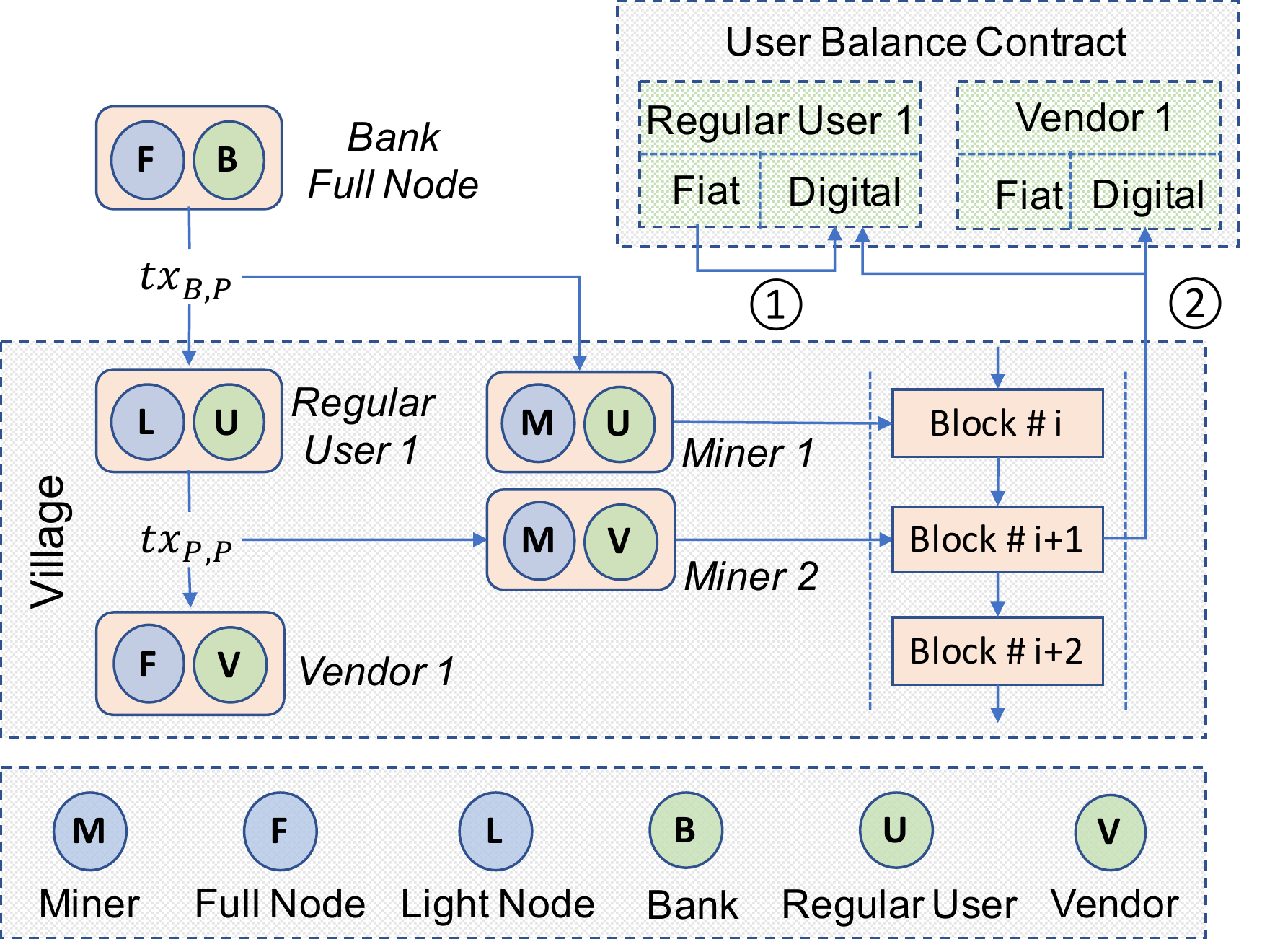}}
	
	\caption{Overview of the proposed system.} \vspace{-4mm}
	\label{fig:sys_overview}
\end{figure}

\figurename~\ref{fig:sys-architecture} graphically illustrates our system design. The bank authorizes a set of selected villagers to act as miners and compensates them by assigning a certain number of Tokens for each valid block they create. Regular users can join the system by running a light node that is deployed as a customized mobile app. All villagers, especially vendors, can host a full node as a payment gateway to keep a complete transaction record if they wish so. The detailed design of the mining network, with the appropriate number of miners and the minimal connectivity requirements for the system to provide the intended service, is described in Section \ref{sec:modelling}. The distinguishing feature of the proposed system is that it enables villagers with Tokens to perform \emph{regular transactions} ($tx_{P,P}$ in \figurename~\ref{fig:tx_flow}) among themselves with all \emph{regular transactions} treated equally, processed and authorised by the blockchain miners irrespective of the bank's presence.  



The bank itself hosts a passive full node to avoid creating forks during its disconnection periods. It deploys a smart contract, i.e. the \emph{user balance contract}, to record user balances in both fiat currency and digital currency, as well as distributing mining rewards. When the connection between the bank and the village is established, the bank synchronizes with other blockchain nodes, updates user balances and processes currency exchange requests. These \emph{currency exchange transactions} ($tx_{B,P}$ in \figurename~\ref{fig:tx_flow}) between the bank and villagers are treated with a higher priority by the blockchain network as they are only committed when the connection is available. The bank also takes appropriate actions in case of malicious behaviors such as suspicious transactions and misbehaving miners. As a result, we have developed the first ever smart contract responsible for the admission control, account management, rewarding miners, and token creation. 

\figurename~\ref{fig:tx_flow} illustrates transaction flows in the system. In order to send a transaction, Regular User 1 first requests to load her digital account with the money in her fiat account.\footnote{A fiat account records the amount of fiat currency a user deposits to the bank.} The bank then issues transaction $tx_{B,P}$, which is picked up by Miner 1 and included in Block \#i. Once $tx_{B,P}$ is confirmed, the code inside the \emph{user balance contract} is triggered and User 1's fiat account and digital account are updated accordingly as shown in Step \textcircled{1}. Next, User 1 sends Vendor 1 a transaction $tx_{P,P}$ to obtain the service, which is collected by Miner 2 and confirmed in Block \#i+1. During the bank's next synchronization, once it reaches Block \#i+1, the smart contract code is triggered again and the digital accounts of both User 1 and Vendor 1 are updated as shown in Step \textcircled{2}. The bank also keeps two accounts of itself, just like other blockchain nodes. If an outside party without Tokens wants to make transactions with people in the village, it first submits a request to the bank node in fiat currency. The bank will then convert the payment to the digital currency locally and redirect it to the target receiver when the connection is available.

\section{System Modelling and Design}
\label{sec:modelling}
We first probabilistically model transaction processing of the local blockchain system as it is a major concern of users. We then model the overall deployment and operational costs from the bank's point of view and use the models to design system properties to enable an efficient and reliable local area payment scheme under the given network constraints. 


\subsection{Assumptions}
We make the following assumptions without loss of generality:
\begin{enumerate}
	\item All individual mining nodes have equal and stable computational power as mining equipment are supplied by the bank to authorised villagers. 
	\item Block size is sufficient to include all transactions for immediate processing as the transaction rates within a village will be considerably lower when compared to a public Ethereum network.
	\item No transaction fees and no rewards for mining stale blocks as rewarding is controlled by the bank using Tokens.
	\item Network bandwidth available within the village is sufficient for all blockchain related data traffic.
\end{enumerate}

\subsection{Modelling Transaction Processing}
\label{subsec:tx_processing}
Regular transaction arrival $r_{t_i}$, where $t_i$ is the transaction initiation time, follows a Poisson distribution as observed in \cite{kasahara2016priority}. We denote the average arrival rate as $\lambda_t$ transactions per second (tps). 
Block generation time $T$ has been shown to be exponentially distributed with probability density function $G(T)=\lambda e^{-\lambda T}$, where $E[T]=\frac{1}{\lambda}$ \cite{decker2013information}. In Ethereum, $E[T]\approx 12s$. Since we assume block size to be large enough (cf. assumption 2), there are no pending transactions as all transactions arriving in the current mining session are included in the next block. Therefore, transaction processing time $t_p = T- t_i$, where $t_i, t_p \in (0, T]$. We further analyze the distribution of block time and transaction processing time under different arrival rates in Section \ref{subsec:numerical}. 
Transaction throughput can also be calculated as $\frac{\sum_{\forall t_i \leq T} r_{t_i}}{T}$, with transaction size being $s_t$ bits, blocks need to accommodate $s_b=\lambda_ts_tE[T]$ bits of transactions on average. 

Then, there are \emph{money exchange transactions} with the bank when it is connected. We model money exchange transaction proccessing similar to that of regular transactions with average arrival rate $\lambda_e$ and size $s_e$.

\subsection{Modelling the System Cost}
We derive cost models from the bank's point of view for capital equipment and operational costs based on metrics in \cite{croman2016scaling}. We omit transaction validation and storage costs in our calculations as they are only a small fraction of mining equipment cost and mining itself \cite{croman2016scaling}. 


\subsubsection{Equipment Cost}
Mining equipment may come in different forms. We use $d_m$ to denote cost of a mining equipment and $l_m$ for total number of miners with a life time of $x_m$ (in years). Then, the averaged equipment cost per year is $C_E=l_m d_m/x_m$.

\subsubsection{Rewarding Cost}
Although the block reward is an incentive for miners, it is an extra cost for the bank. We use $R$ to denote the block reward, meaning that bank has to pay $R$ Tokens for each valid block. We calculate the rewarding cost as $C_R = nR$, where $n$ represents the number of blocks.

\subsubsection{Network Resources}
Since we assume ideal network conditions (cf. assumption 4) locally, we only model the network resource usage of the backhaul network. When connected, the bank synchronizes past transactions and processes money exchange requests. We define one service period as $T_B=T_C+T_U$, in which $T_C$ and $T_U$ stand for the cumulative total duration of connected and disconnected sessions respectively during the service period. 
We denote the backhaul network bandwidth requirement as $BW$ and cost as $C_{BW}$ in Token per bit. Then, the network connectivity cost would be $C_{N}=C_{BW}T_CBW$ during one service period. 

\subsubsection{System Cost}
To sum up, the overall system cost bank has to afford is the sum of all the aforementioned costs. Since cost components are defined with different time units, we calculate the overall cost within time period $T_0$, which equals $x_y$ years, $x_b$ blocks and $x_s$ service periods as:
\begin{equation} \label{eq:final_cost}
\begin{split}
C_{All} &=C_{E}x_y+C_{R}+C_{N}x_s \\
&=l_m\frac{d_m}{x_m}x_y+ Rx_b + C_{BW}T_CBWx_s.
\end{split}
\end{equation}


\subsection{Mining Network Design}
\label{subsec:design_constraints}
We now find ranges of multiple mining network parameters including number of miners and their connectivity to ensure reliability and security. 

\subsubsection{Miner Outages}
Miners are incentivized to work, but they may join or leave the network spontaneously (churn) \cite{heilman2015eclipse}. We denote total number of online miners as $l_{on}$, $l_{on}\leq l_m$. Probability of one miner going offline in each time slot is $p_1, p_2, p_3, \cdots, p_{l_m}$. Hence the probability that $X=k$ miners are offline in the same time slot can be represented by a Poisson Binomial Distribution:
\begin{equation}
P_r(X=k)=\sum_{A\in F_k} \prod_{i\in A}p_i \prod_{j\in A^c}(1-p_j),
\end{equation}
where Set $F_k$ contains all subsets of k integers that can be selected from $\left\{1,2,3,\cdots,l_m\right\}$, and $A$ and $A^c$ are complementary. If all miners have the same chance of going offline, i.e. $p_1=p_2=p_3=\cdots=p_{l_m}=p_d$, the expectation of $l_{on}$ is:
\begin{equation}
E[l_{on}]=l_m-E[X]=l_m (1-p_d),
\end{equation}
or equivalently,
\begin{equation} \label{eq:miner_with_outages}
l_m = \frac{E[X]}{p_d} = \frac{E[l_{on}]}{1-p_d}.
\end{equation}
We demonstrate how network churn affects the blockchain system performance in Section \ref{subsec:experiments}.

\subsubsection{Mining Reward}
Bank determines the mining reward $R$ according to the status of local economy and its budget. It becomes harder for each individual miner to find a valid block when there are more competitors (cf. assumption 1). As the expected reward reduces, mining becomes less profitable. We here calculate the maximum number of miners allowed to keep them incentivized. We denote the mining cost as $\eta$ per hash, similar to the one in \cite{rizun2015transaction}. With individual hash rate being $h$, each miner's operational cost per second is $\eta h$. The expected hash cost of the entire network per second $C_H$ when all miner are online (worst case) is $C_H=l_m\eta h$. Since miners are equal, their expected revenue per mining round is hence $R'=R/l_m$. We do not consider rewards provided for stale blocks (cf. assumption 3). We derive the expected profit per mined block for each individual as:
\begin{equation}\label{eq:individual_profit}
\Pi =R'-\frac{C_H T}{l_m}=\frac{R}{l_m}-\eta hT.
\end{equation}
Clearly, for mining to be profitable, $\Pi$ should be greater than 0, i.e. $l_m$ and $R$ should satisfy:
\begin{equation}
l_m<\frac{R}{\eta hT}.
\end{equation}
%

\subsubsection{Minimal Connectivity Requirement}
\label{subsec:min-connections}
A key factor of blockchains is the synchronization across the network, i.e. every node should obtain a copy of the ledger. Individual miners may want to reduce their data usage by connecting to fewer other nodes, but with security considerations all miners are encouraged to connect to as many other peers as possible. We here provide an intuitive picture on the minimal direct connections required for efficient information propagation across the network. For the simplicity of analysis, we assume on average every miner maintains connections with $l_c$ other nodes, $0<l_c\leq l_m$. Once Miner A receives a transaction or a block, she sends it to all the directly connected nodes. The same procedure repeats until the message reaches every node in the network. We define information propagation from one network node to another as a hop. Restricting maximum number of hops to $k$, we obtain: 
\begin{equation} \label{eq:min_connections}
\begin{split}
& 1+l_c+l_c(l_c-1)+\cdots +l_c(l_c-1)^{k-1} \\
& = 1+l_c\sum_{i=0}^{k-1}(l_c-1)^{i} \geq l_m.
\end{split}
\end{equation}
Solving Equation \ref{eq:min_connections} under given $k$ and $l_m$ (not $l_{on}$ as eventually all miners will store the blockchain), we obtain $\gamma$, fraction of nodes that one should directly connect to as: 
\begin{equation}
\gamma \geq \frac{{min}\left(l_c\right)}{l_m}.
\end{equation}

\subsection{Summary}
Although the bank has every intention to have fewer mining nodes to reduce the equipment cost, there should be sufficient miners to keep the fairness of the system and meet the security requirements. The bank should also combine real-world churn rate data and equipment costs to determine the number of mining nodes needed and device type according to particular use cases. Furthermore, all villagers, especially miners, are recommended to increase their connectivity for a better synchronization even though this leads to higher bandwidth usage. 
\section{Evaluation}
\label{sec:evaluation}
In this section, we first validate our transaction processing model by comparing it to a real deployment of a private Ethereum blockchain and make predictions about the system behaviour based on the validated model. We also dimension the local blockchain network design by identifying the minimal average connection required for each mining node. Finally, we evaluate the blockchain performance under various \emph{Network Delays} and \emph{Network Churns} on the experimental deployment.

\subsection{Experimental Environment}
\label{sec:setup}
We have deployed a private Ethereum blockchain residing on multiple virtual machines running Ubuntu Linux v16.04 in an Openstack environment.\footnote{\url{https://www.openstack.org}} Each virtual machine is given 1 virtual CPU core, 2 GB of memory, and 10 GB of persistent storage to meet the minimum hardware requirement for running Ethereum. Elastic Search\footnote{\url{https://www.elastic.co}} was used to store block-related information. The network behaviour was monitored using the Python Web3 Library.\footnote{\url{https://web3py.readthedocs.io/en/latest/index.html}}


All virtual machines are linked together in a low-latency local network that can be customized on demand.
Our setup peers every node using a single Gigabit Ethernet switch and forms a star topology similar to the cellular network in \figurename~\ref{fig:sys-architecture}.
With this setup, a communication round-trip time between every two nodes is less than 1 ms on average. 
We employ Linux traffic control to introduce delays to emulate high latency in mobile networks and configure the Linux kernel firewall with Iptables\footnote{\url{https://help.ubuntu.com/community/IptablesHowTo}} to emulate churns.

We have used Geth v1.6.4~\footnote{\url{https://github.com/ethereum/go-ethereum}} for all of our empirical evaluations. Before the actual experimentation, we let the system stabilize for a few hours to obtain appropriate parameters of the genesis block in later runs. Based on our observation, we set the initial \emph{nonce value}, \emph{gas limit} and \emph{difficulty} to 0x42, 0x08000000, 0x400000 respectively to make the system stabilize quickly.
To form a fully connected P2P blockchain network, we disabled the auto-discovery feature supported by Geth and configured the overlay network manually.
Finally, if not mentioned otherwise, all experiments involved 10 mining nodes and 10 light nodes.


\subsection{Model Validation and Numerical Sensitivity Analysis}
\label{subsec:numerical}

\subsubsection{Model Validation}
\label{subsubsec:model-validation}
\begin{figure}[!htbp]
	\centering
	\resizebox{0.5\textwidth}{!}{%
		\subfloat[\scriptsize{Block time.}]{%
			\includegraphics[scale=0.4]{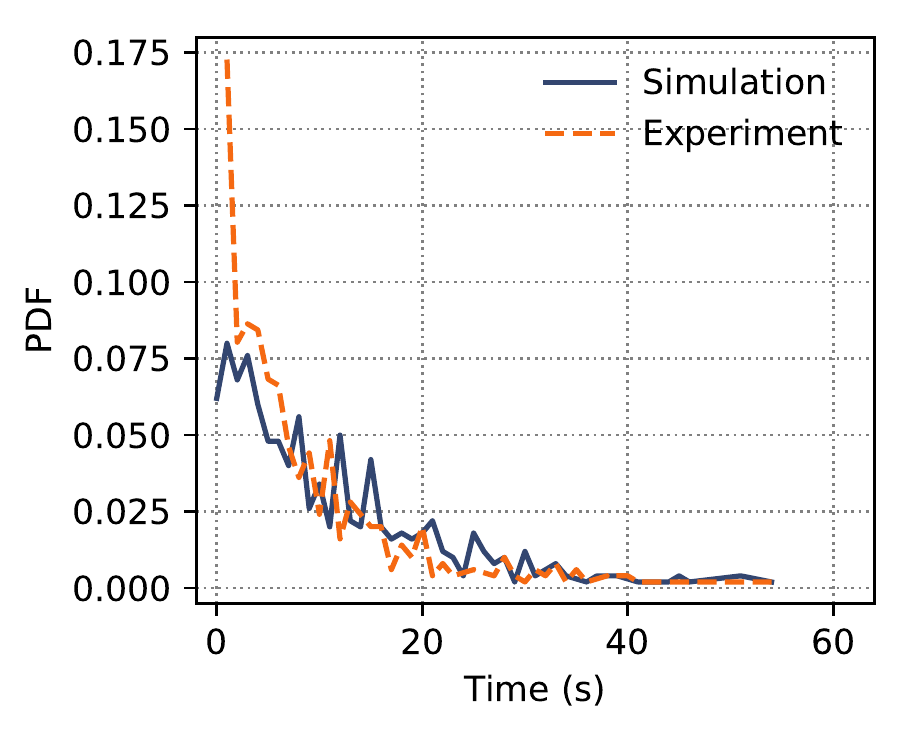} 
			\label{fig:blk-sim-exp-comp}
		}
		\subfloat[\scriptsize{Transaction processing time.}]{%
			\includegraphics[scale=0.4]{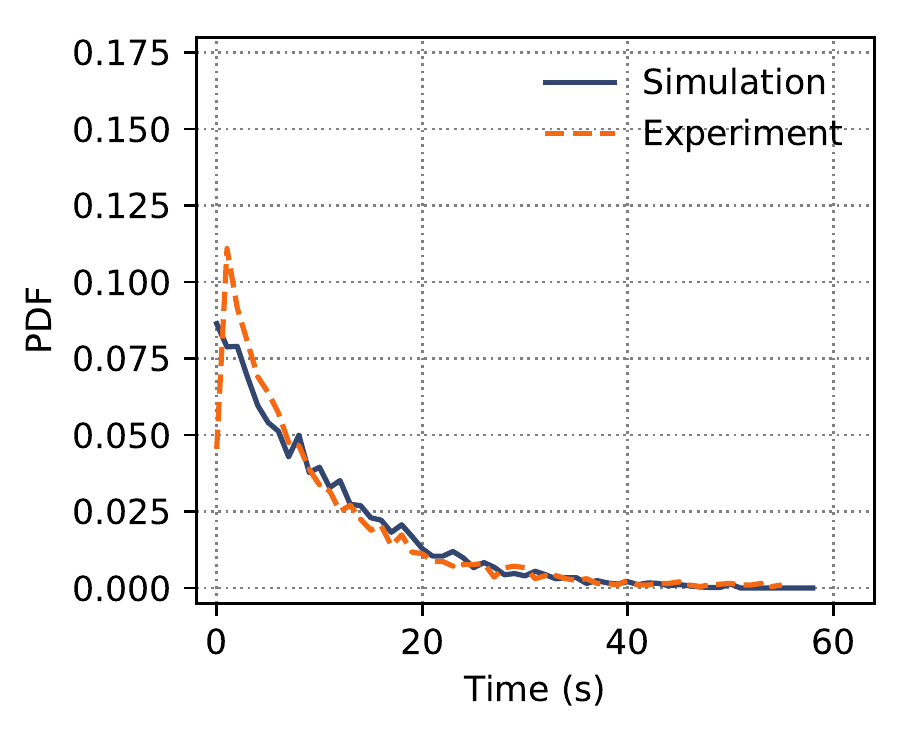} 
			\label{fig:tx-sim-exp-comp}
		}
	}
	\caption{Model validation in terms of block generation and transaction processing at 1tps.} \vspace{-5mm}
	\label{fig:sim-exp-comparison}
\end{figure}

To validate our model in Section \ref{subsec:tx_processing}, we emulated a transaction processing scenario by issuing transactions at 1 tps (overall rate) from multiple clients to randomly selected addresses. We sort block timestamps in ascending order and calculate block generation time as the difference between two adjacent timestamps. To obtain transaction processing time, we record the timestamp upon initiation of a transaction and extract the inclusion timestamp from the transaction receipt. \figurename~\ref{fig:blk-sim-exp-comp} and \figurename~\ref{fig:tx-sim-exp-comp} present the simulation and emulation results of 500 consecutive blocks on block time and transaction processing time respectively. Both illustrate a strong agreement between modelling and experimental measurements. Additionally, an inter-comparison between \figurename~\ref{fig:blk-sim-exp-comp} and \figurename~\ref{fig:tx-sim-exp-comp} confirms that transaction processing and block generation are highly correlated.
\vspace{-5mm}

\subsubsection{Impact of Transaction Arrival Rate}
\label{subsubsec:impact-of-tx-rate}
We then intended to empirically study the impact of transaction rate on transaction processing and benchmark the local blockchain performance. However, when increasing the arrival rate, we hit the transaction sending limit before the processing limit due to implementation bugs in the Ethereum version we used. Indeed, with more than 3 tps, the transaction generator stopped working due to connection errors as the algorithm thinks some nodes are flooding the system with too many requests. Despite the transaction generator worked smoothly under lower arrival rates, in the end only 4246 out of 17265 transactions were mined at 2 tps. We hence decided to analytically study the system behaviour under multiple transaction rates.

\begin{figure}
	\resizebox{0.5\textwidth}{!}{%
		\subfloat[\scriptsize{Block time.}]{%
			\includegraphics[scale=0.4]{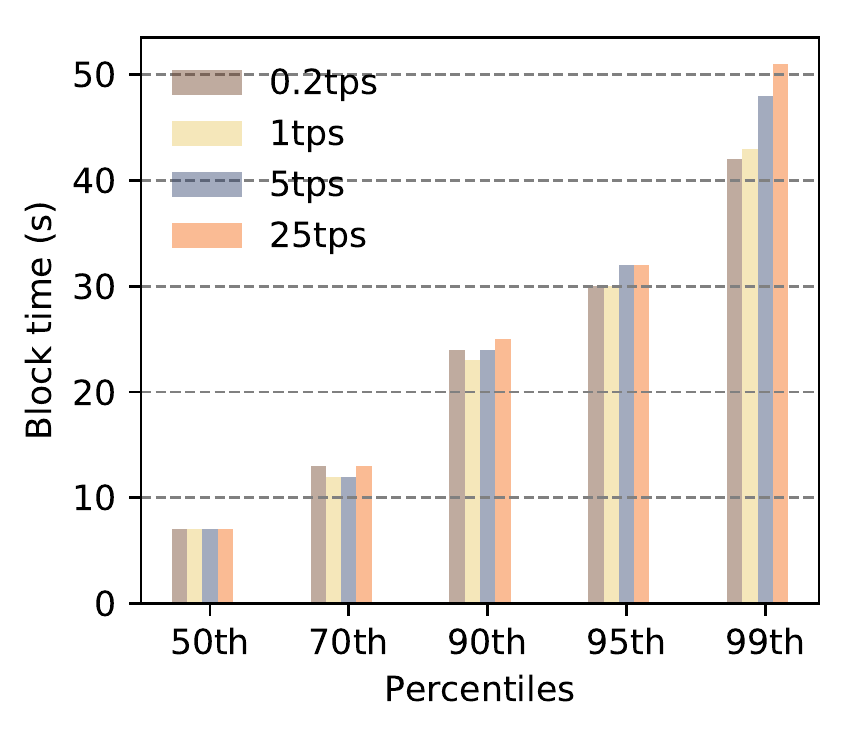} 
			\label{fig:sim_blocktime_percentiles}
		}
		\subfloat[\scriptsize{Transaction processing time.}]{%
			\includegraphics[scale=0.4]{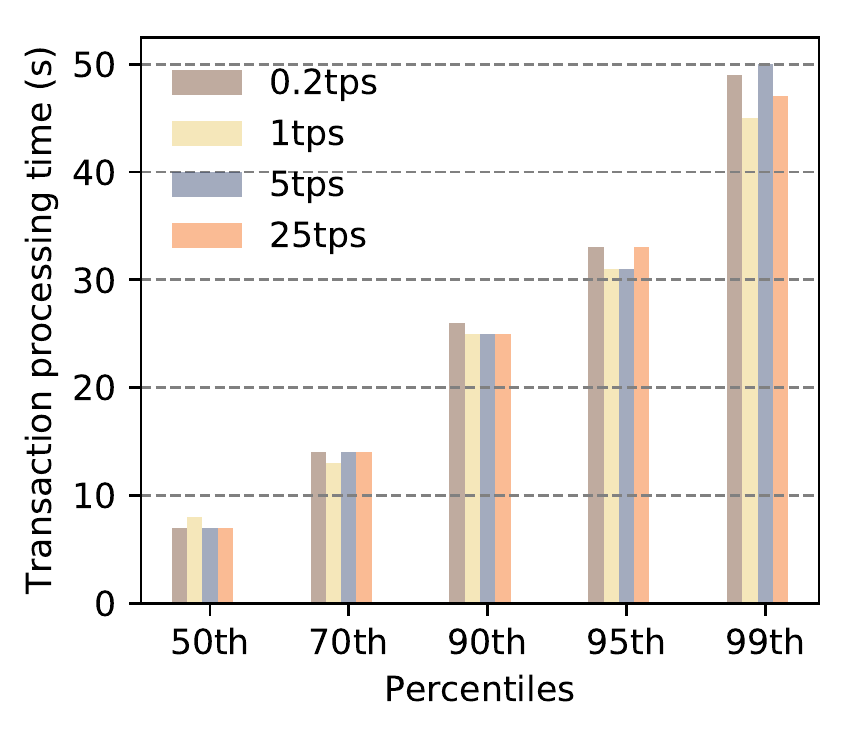} 
			\label{fig:sim_txtime_percentiles}
		} 
	}
	\caption{Simulation with multiple transaction rates.} \vspace{-4mm}
	\label{fig:sim_tx_processing}
\end{figure}

Using the validated model in Part~\ref{subsubsec:model-validation}, we further generated transactions at various rates including 0.2 tps, 1 tps, 5 tps, and 25 tps in our simulation. \figurename~\ref{fig:sim_blocktime_percentiles} and \figurename~\ref{fig:sim_txtime_percentiles} compares the 50th, 70th, 90th, 95th and 99th percentiles of block time and transaction processing time under all the simulated arrival rates. These results confirm that with a sufficient block size, block creation time and transaction processing time are not affected by the arrival rate.

\subsubsection{Minimal Connection Requirement}
\begin{figure}[!htbp] 
	\centering
	\resizebox{0.5\textwidth}{!}{%
		\subfloat[\scriptsize{\# of connections.}]{%
			\includegraphics[scale=0.4]{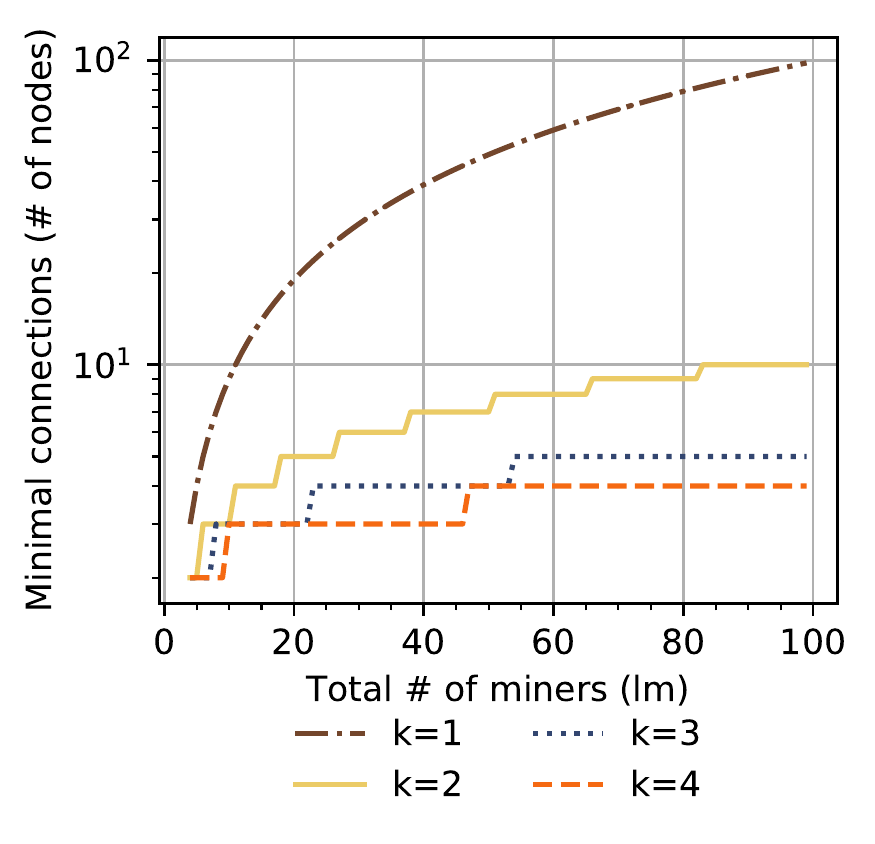}
			\label{fig:min_connect_number}
		}
		\subfloat[\scriptsize{Fraction of the network.}]{%
			\includegraphics[scale=0.4]{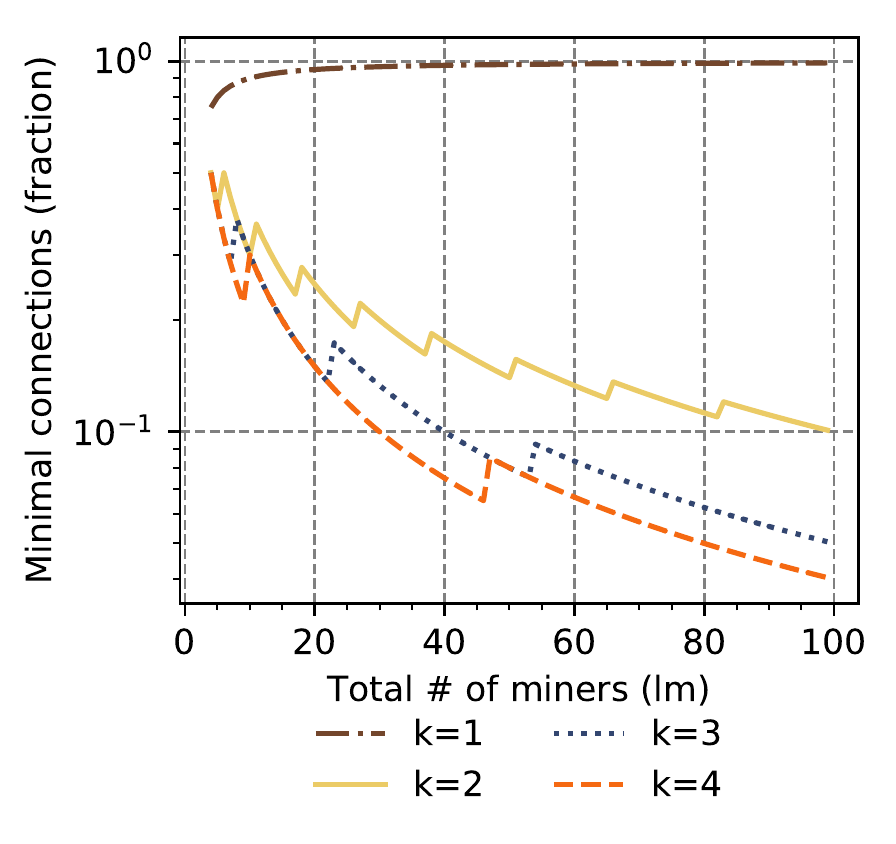}
			\label{fig:min_connect_frac}
		} 
	}
	\caption{Minimal connection requirement.} \vspace{-5mm}
	\label{fig:min_conectivities}
\end{figure}

Based on the analysis presented in Part~\ref{subsec:min-connections}, we obtain the minimal connection requirement when varying number of hops $k$ from 1 to 4 in (\ref{eq:min_connections}) with a total number of miners $l_m$ increasing from 4 to 100. \figurename~\ref{fig:min_conectivities} displays the results in the number of nodes and fraction of the network. As the left-hand side of (\ref{eq:min_connections}) increases exponentially with $l_c$, minimum $l_c$ that satisfies the equation may remain unchanged for a range of $l_m$, which results in steps or spikes on the curves. As shown in \figurename~\ref{fig:min_connect_number}, when $k=2$, with a total of 100 miners, each individual needs to maintain at least 10 connections. When $k=3$, the minimum number of connections required reduces to 4. Generally speaking, if only one hop is allowed for information propagation, all miners need to connect with all other miners. While if more hops are allowed, miners can limit their connections to reasonably small values to save the bandwidth usage. This indicates the system's ability to scale without miners exhausting their network resources.

\subsection{Effect of Network Disturbances} 
\label{subsec:experiments}
To further evaluate the performance of the local blockchain network, we investigate its behavior under disturbances such as \emph{Network Delays} and \emph{Network Churns}. Since it was impossible to obtain the whole picture from transaction processing (cf. Part~\ref{subsubsec:impact-of-tx-rate}), we have considered block generation time to be an alternative evaluation metric (cf. Part~\ref{subsubsec:model-validation}). 
\subsubsection{The (non)Impact of Network Delays}
\begin{figure}[!htbp]
	\centering
	\subfloat[\scriptsize{Block time percentiles.}]{%
		\includegraphics[width=.8\columnwidth]{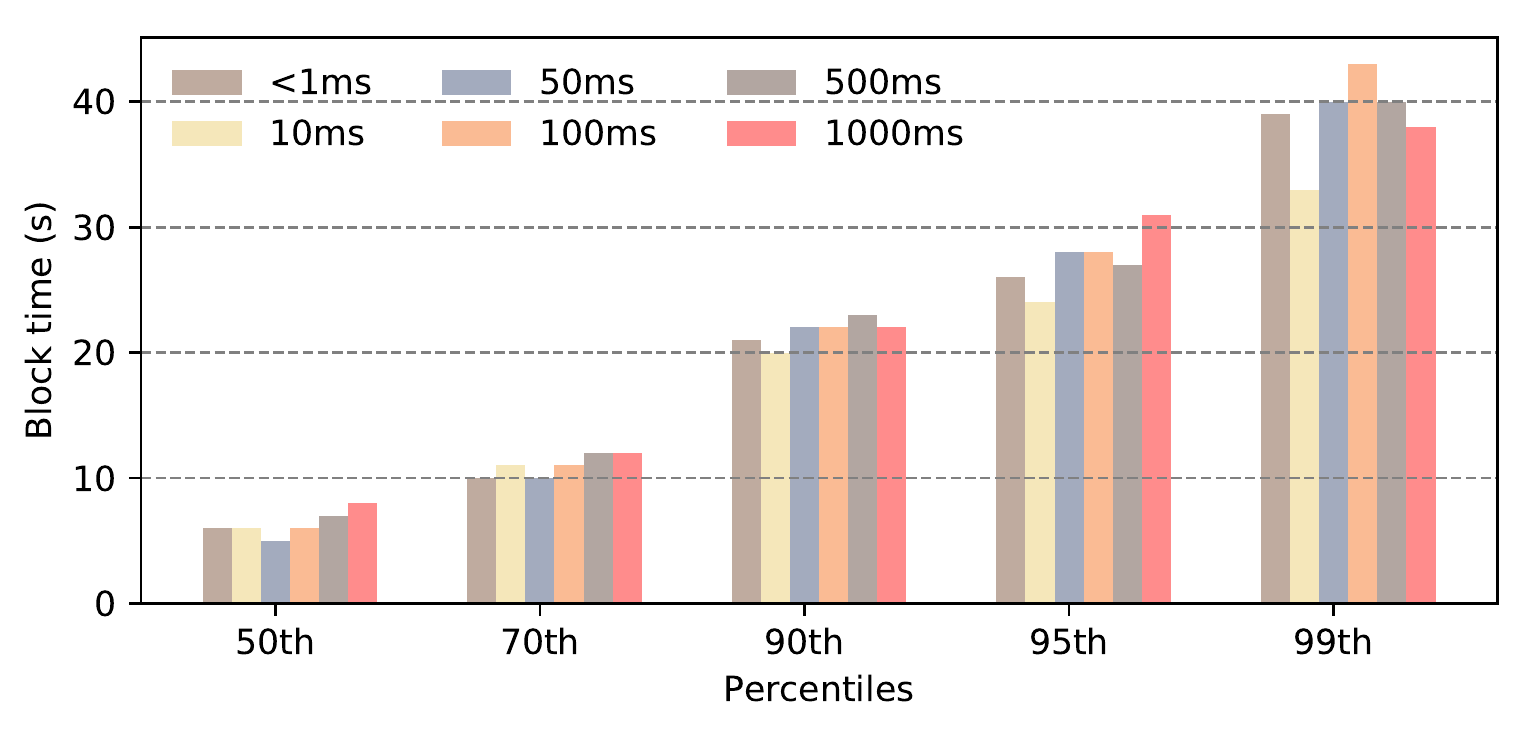}
		\label{fig:bars-btpercentile-withdelay}
	}
	
	\subfloat[\scriptsize{PoW difficulty levels.}]{%
		\includegraphics[width=.8\columnwidth]{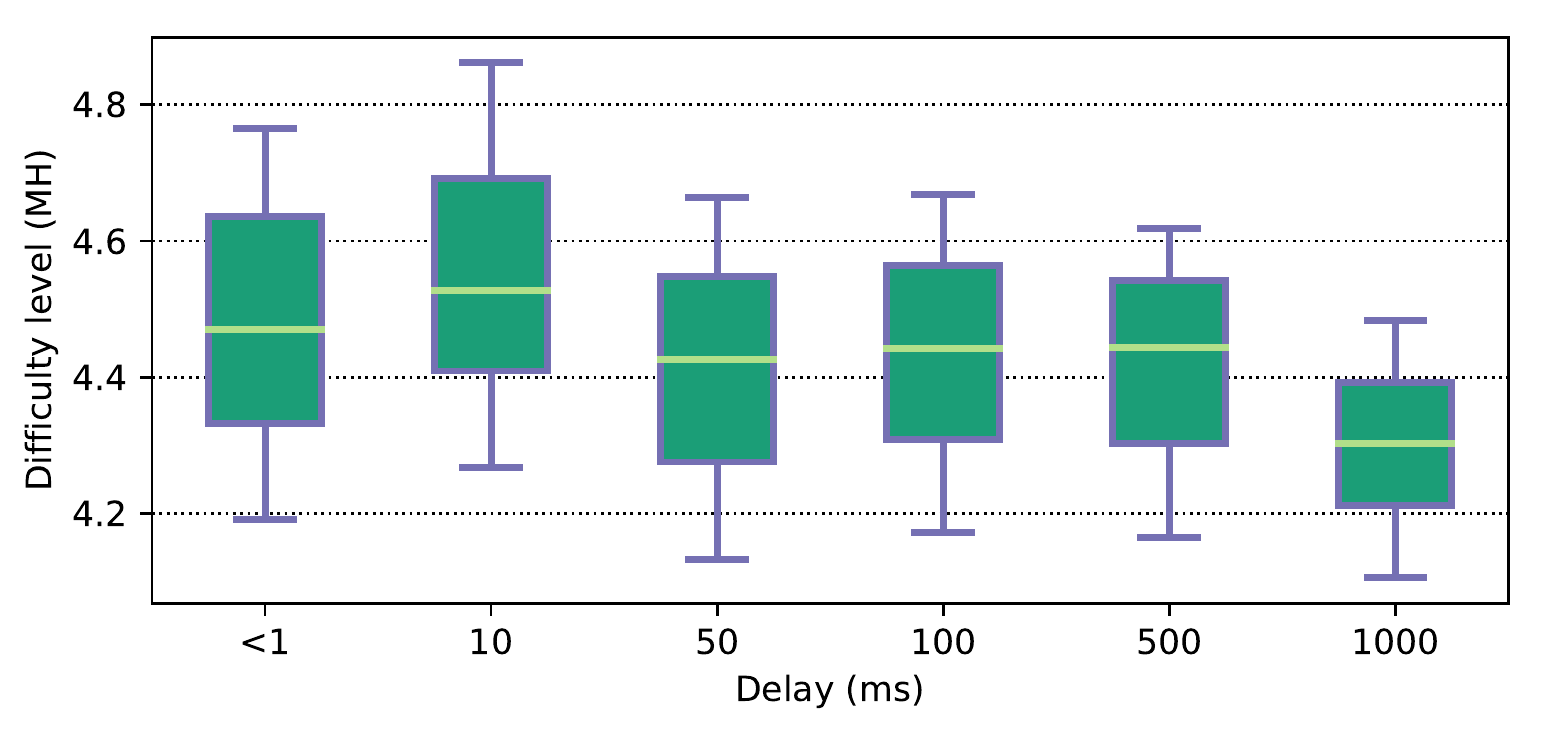}
		\label{fig:bp-difflv-withdelay}
	} 
	\caption{System behaviour with network delays.} 
	\vspace{-5mm}
	\label{fig:delays}
\end{figure}

In order to understand the stability of our proposal when inter-node delay increases, we introduced various delays including 0, 10 ms, 50 ms, 100 ms, 500 ms, 1000 ms per connectivity channel.\footnote{Today's mobile network delays can hardly go beyond 1 second: \url{https://hpbn.co/mobile-networks/}} \figurename~\ref{fig:bars-btpercentile-withdelay} shows the 50th, 70th, 90th, 95th, 99th percentiles of block time, from which we can see block time stays stable even with delays reaching 1 second. We ascribe this to PoW difficulty adjustment where network delay causes time difference between two adjacent blocks to increase, the internal algorithm reduces difficulty level to achieve a shorter block time. \figurename~\ref{fig:bp-difflv-withdelay} illustrates block difficulty under multiple network delays. A decreasing trend can be observed in the difficulty level when delay increases. This behaviour helps to maintain the transaction processing speed, but is ``unhealthy'' because with a fixed total network hash rate, a lower difficulty level leads to more stale blocks and inconsistencies.

\subsubsection{Network Churns} 

\begin{figure}[!htbp]
	\centering
	\includegraphics[width=.9\columnwidth]{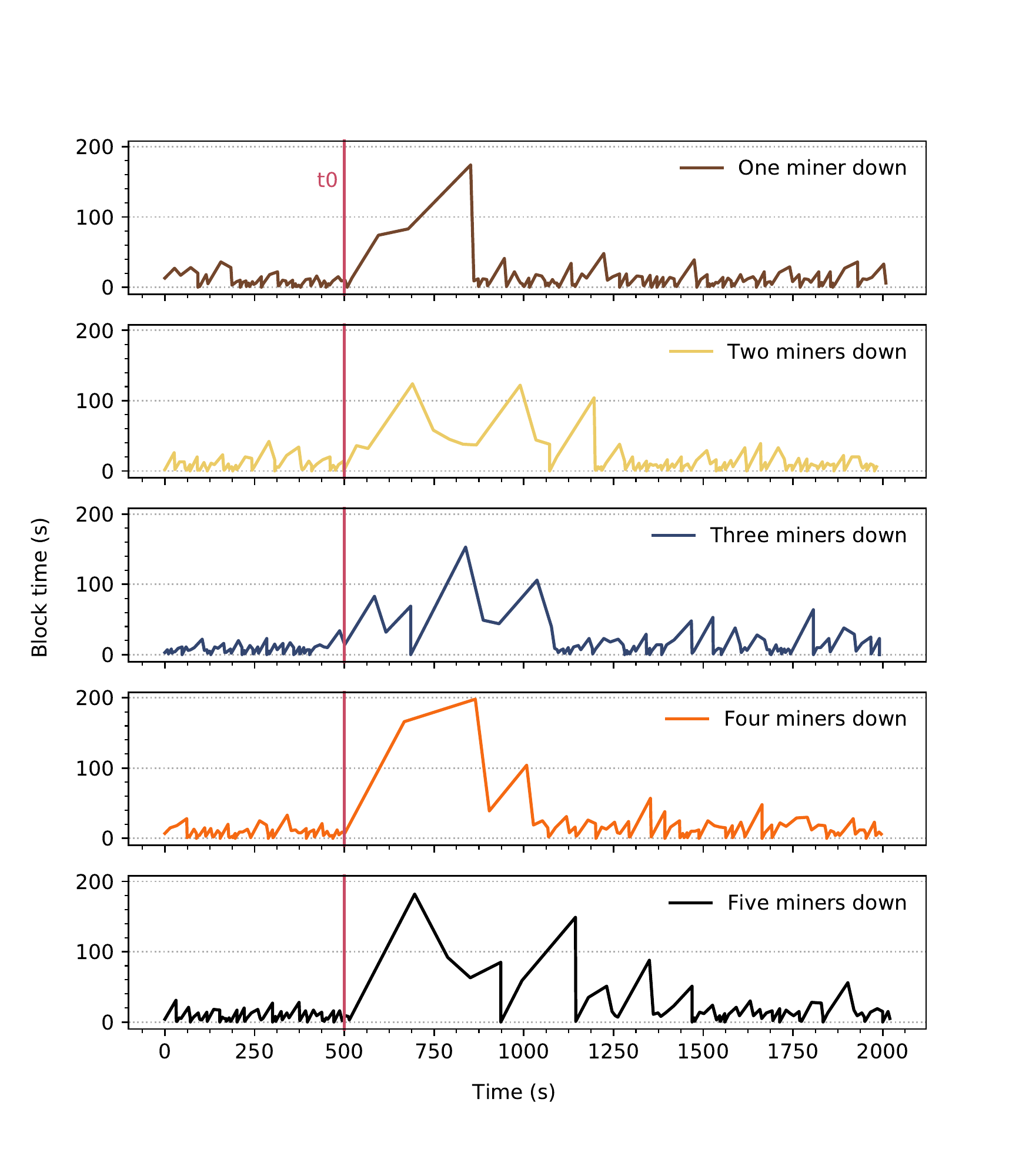} \vspace{-5mm}
	\caption{Transient response to mining nodes going offline.} \vspace{-5mm}
	\label{fig:one_off_churns}
\end{figure}

\paragraph{Transient Response} 
One of the main characteristic of P2P networks concerns the churn rate of their nodes. We emulated a scenario in which some miners go offline at time $t_0$ and observed variation in block creation time. As shown in \figurename~\ref{fig:one_off_churns}, when some miners become offline, block time experiences a significant increase followed by a ``resolving period'' during which PoW difficulty is adjusted in response to changes in network hash rate. To find the end point of the ``resolving period'', we select the timestamp from which the absolute variation is less than 5\% for the next 3 consecutive points. Duration of ``resolving period'' is measured as 361 s, 781 s, 577 s, 527 s, 969 s respectively. Overall, it increases with the number of offline nodes. While the network recovers, we have also observed the behaviour of ``backwards mining'', meaning that newer blocks sometimes have smaller block number than older ones. The disagreement is resolved with the stabilization of the network. 

\begin{figure}[!htbp]
	\centering
	\includegraphics[width=.8\columnwidth]{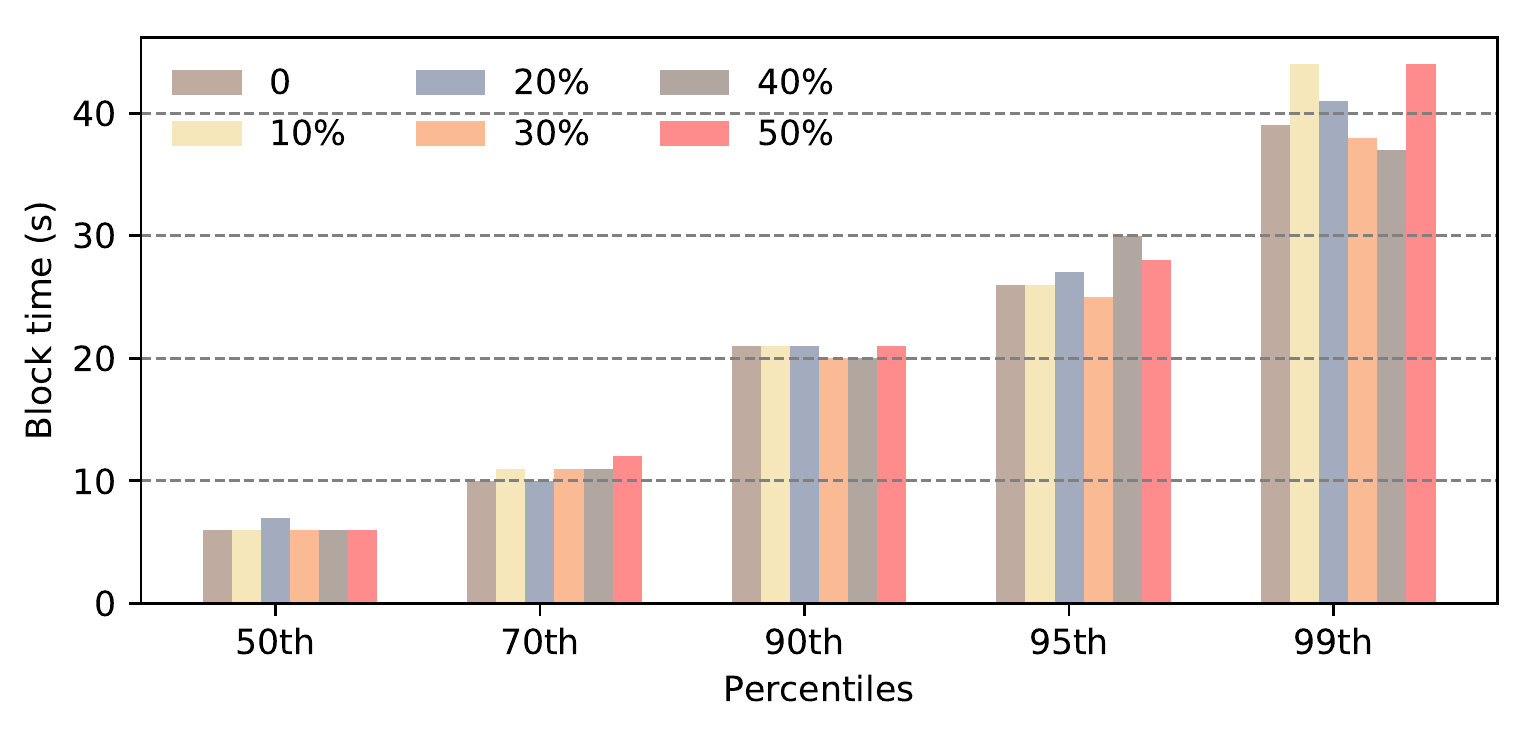}
	\caption{System steady-state behaviour with network churns.} \vspace{-5mm}
	\label{fig:bars-btpercentile-withchurn}
\end{figure}

\paragraph{Changes Over Time}
We then tested system performance when each miner has a churn rate of 10\%, 20\%, 30\%, 40\% and 50\%. We ran each set of experiment for approximately 2 hours and changed miner status (online or offline) every 20 minutes. \figurename~\ref{fig:bars-btpercentile-withchurn} shows the 50th, 70th, 90th, 95th, 99th percentiles of block generation time, from which we can see block time does not vary much across all churn rates. In other words, number of miners do not affect speed of block generation or transaction processing.
%

\subsubsection{Summary}
Despite being unable to test the processing limitation due to Ethereum implementation flaws and obtain all transactional data experimentally, we managed to study the effect of network disturbances from block time. Difference between empty blocks and blocks with transactions mainly lies in block size, hence the additional transmission delay which is expected to have similar impacts to network delays. Results show that delays and churns may cause block generation time to change especially in the transient state, however, due to the difficulty adjustment they do not necessarily slow down block generation in the long run. Overall, the main observed drawback of the local blockchain is that disturbances may cause more temporary inconsistencies.

\section{Implementation}
\label{sec:implementation}
In this section, we demonstrate the feasibility of the system design with a prototype implementation containing a private Ethereum blockchain and an intermittently connected bank full node. We tested the synchronization delay of the bank node when varying the bandwidth and disconnection duration. We also measured the usage of data, CPU, and battery of the mobile wallet app using DDMS \footnote{\url{https://developer.android.com/studio/profile/ddms.html}} and Battery Historian. \footnote{\url{https://developer.android.com/studio/profile/battery-historian.html}} 

\subsection{Setup}
\begin{figure}[!ht]
	\centering
	\includegraphics[width=.8\columnwidth]{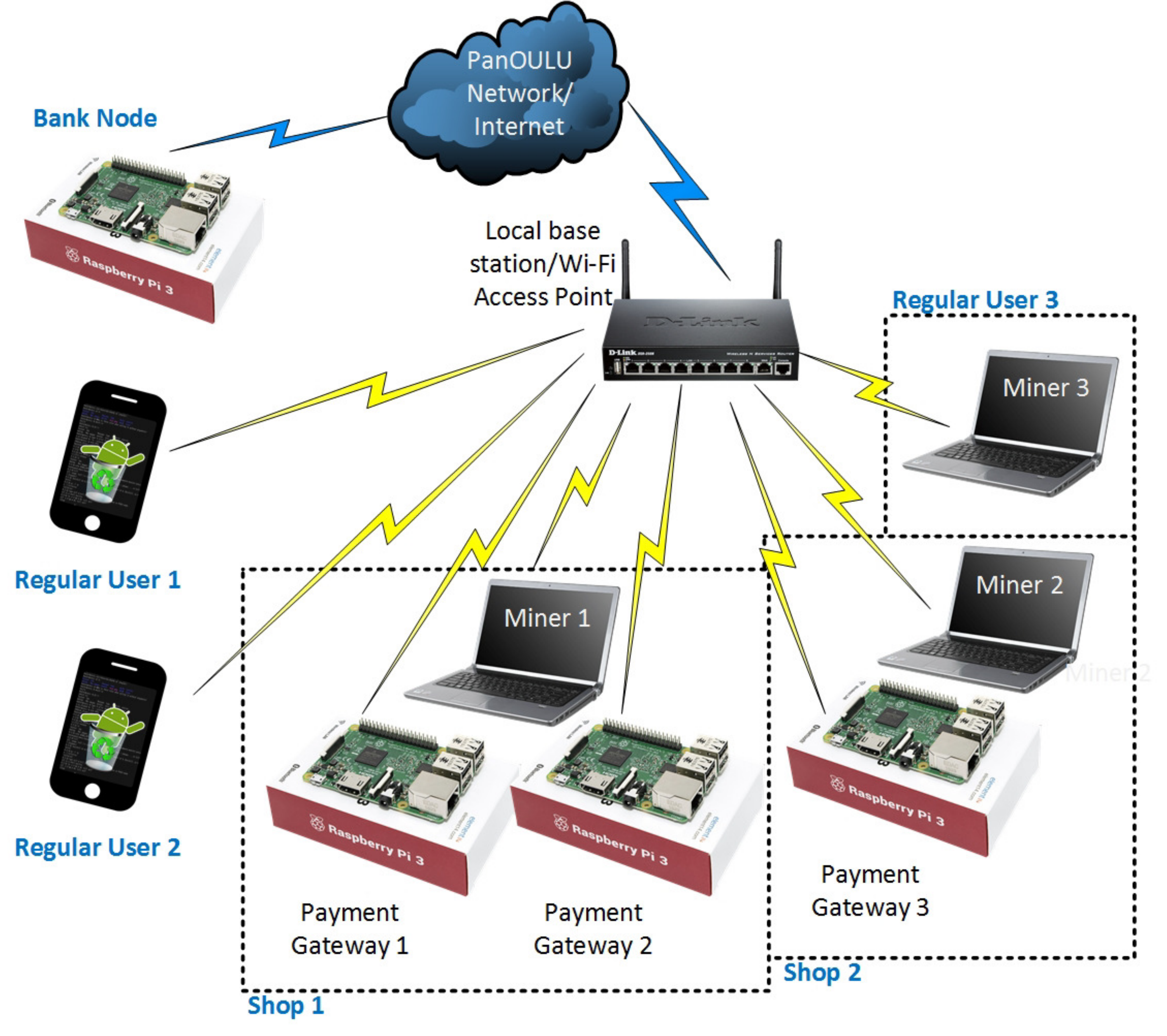}
	\caption{Prototype implementation.} \vspace{-5mm}
	\label{fig:prototype}
\end{figure}

Figure \ref{fig:prototype} illustrates the setup with a bank full node, 2 shops, and 3 regular users. Each shop runs a mining node and in addition, Shop 1 owns two full nodes and Shop 2 owns one full node. Regular User 3 operates a miner while the other two regular users are light mobile clients. We summarize the device types, capabilities and Geth versions in Table \ref{tb:device-model-geth}. 
\vspace{-3mm}
\begin{table}[!ht]
	\centering
	\caption{Devices and their capabilities.} \vspace{-2mm}
	\begin{tabular}{>{\centering\arraybackslash}m{1.4cm}  >{\centering\arraybackslash}m{2.1cm}  >{\centering\arraybackslash}m{1.8cm}  >{\centering\arraybackslash}m{1.8cm} } \specialrule{.12em}{1em}{0em}
		 & {\bf Miner} & {\bf Full node (including bank)} & {\bf Light node} \\ \hline \hline
		{\bf Device} & Dell Latitude 6430u & Raspberry Pi 3 & Google Pixel XL\\ \hline
		{\bf \# of nodes} & 3 & 4 & 2 \\ \hline
		{\bf OS} & Ubuntu 17.04  & Raspbian Jessie & Android v8.0.0 \\ \hline
		{\bf RAM} & 4GB & 1GB & 4GB \\ \hline
		{\bf Geth} & v1.6.5 & v1.6.5 & v1.6.7 \\ \hline
		\specialrule{.12em}{0em}{0em}
	\end{tabular}
	\label{tb:device-model-geth}
\end{table}

We used a D-Link DSR-250N Wi-Fi router connecting to the Oulu public WAN network (PanOULU) \cite{ojala2011panoulu} to represent the community base station. We used an IEEE 802.11n Wi-Fi network to interconnect the above nodes except the bank node which is directly connected to the PanOULU network and periodically joins the blockchain via the Wi-Fi router's public network interface. We configured the bank node's backhaul bandwidth via the router's WAN port. We used the auto-discovery protocol of Geth on all nodes. 

We developed a smart contract as the wallet app in Solidity v0.4.12 to create fiat and digital user accounts as well as a special Token to perform functions including token supply, token conversion, token transferring and miner rewarding as described in Section~\ref{sec:sys_architecture}. 



\subsection{Experiments}
\subsubsection{Bank Node Synchronization Delay}
We measured bank synchronization delay (i.e. time taken to update the bank node) while increasing the backhaul bandwidth from 128 Kbps to 10 Mbps. We maintained a constant transaction rate of 2 tps and varied the disconnection time from 1 min to 10 min. \figurename~\ref{fig:bank-sync-delay} illustrates the averaged synchronization delay and variation over 10 runs of each scenario, where synchronization delay is roughly proportional to the disconnection duration under each a fixed backhaul bandwidth.

\subsubsection{Mobile Wallet Performance}
\begin{figure}[!htbp]
	\centering
	\resizebox{0.5\textwidth}{!}{%
		\subfloat[\scriptsize{Bank node synchronization delay.}]{%
			\includegraphics[scale=0.4]{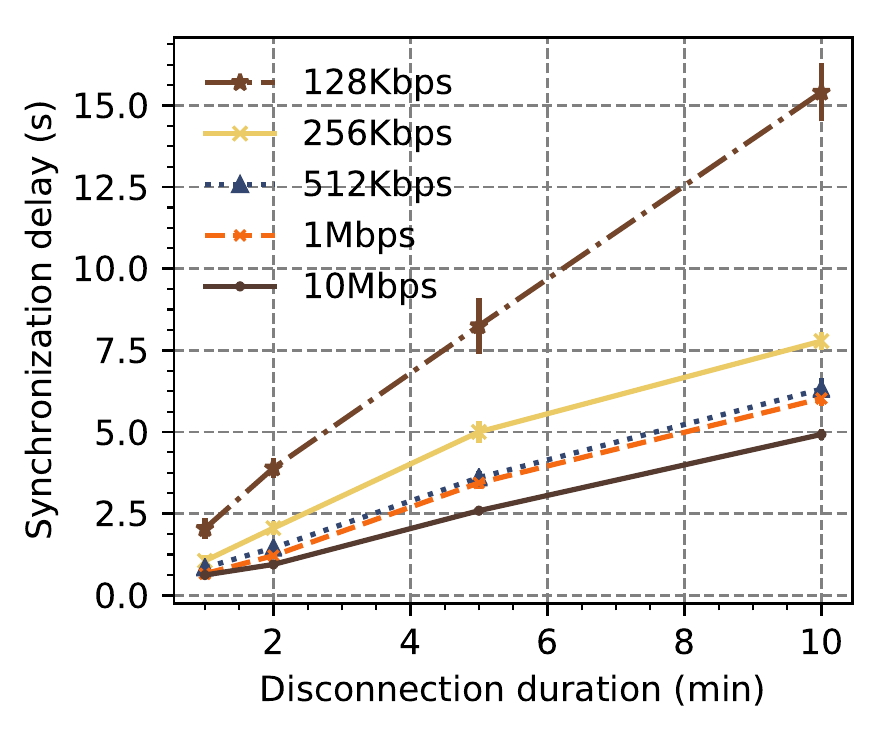} 
			\label{fig:bank-sync-delay}
		} 
		\subfloat[\scriptsize{Mobile wallet data usage.}]{%
			\includegraphics[scale=0.4]{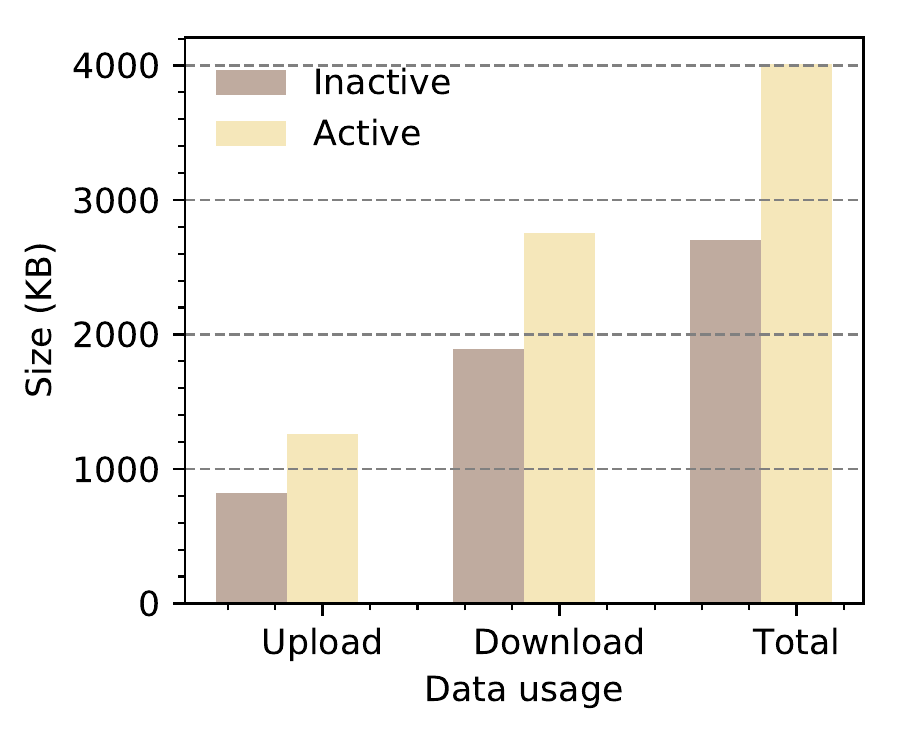} 
			\label{fig:user-data-usage}
		}
	}
	\caption{Test results on prototype implementation.} \vspace{-4mm}
	\label{fig:PTExperiment1}
\end{figure}

We then measured data, CPU and battery used by the mobile wallet in its idle state (i.e. \emph{Inactive} mode) and when it sends one transaction per minute (i.e. \emph{Active} mode) for a total of one hour. Note that without sending any transactions, the mobile light client continuously synchronizes with the network and downloads block headers. As shown in \figurename~\ref{fig:user-data-usage}, our wallet app uses only 2.71 MB in total in its idle state, and an additional 21.65 KB to perform one transaction per minute. The overall CPU usage increased from 51.9 s to 127.7 s within the one-hour period. When it comes to power consumption, the wallet used 3.2 mAh or 0.04\% of the mobile phone's total battery usage in its idle state and 19.8 mAh or 0.11\% when sending transactions. These results confirm our wallet app requires low bandwidth, CPU processing and power for its operations compared to available cryptocurrency clients \cite{bandwidth_bitcoin}.
\vspace{-4mm}



\section{Related Work}
\label{sec:related}
We categorize related work into i) Pervasive, Delay-Tolerant Networks, ii) Micro-Payment Systems, and iii) Blockchain Cost Models.

\textbf{i) Pervasive, Delay-Tolerant Networks:} Blattman et al. \cite{blattman2002assessing} found delay-tolerant networks are sufficient for digital services to meet the needs of most rural communities. Pentland et al. \cite{Pentland2004} later developed DakNet that uses a pervasive mobile coverage to enable asynchronous digital services in India and northern Cambodia. DakNet is remarkably low-cost and well-received by local users and is more accessible than a centralised, community telephone. The success of DakNet has proven that decentralization is an effective way to deliver service to remote regions.

\textbf{ii) Micro-Payment Systems:} A number of micro-payment schemes were proposed in the mid-to-late 1990s. Some of them leverage SMS or USSD of the cellular networks, for instance, the BAAC in Thailand~\cite{fitchett1999bank} and the M-Pesa Service in Kenya~\cite{mas2010mobile}. However, SMS messages are easily spoofable and hence require additional user verifications for security, and USSD could be affected by session time-outs. Other micro-banking systems are the early-generation token payment platforms secured by time-lock puzzles, e.g. PayWord and MicroMint \cite{rivest1996payword}. IBM also proposed a Micro-iKP protocol for frequent micro-payments \cite{hauser1996micro} using ``coupons" sent between players and verified by the bank. This scheme however, increases the communication cost. Bitcoin \cite{nakamoto2008bitcoin}, the first successful P2P cryptocurrency, was established in 2009 and has received great attention. Other alternative coins such as Ethereum \cite{wood2014ethereum} and Dogecoin were later developed. Although blockchain was originally designed to operate without centralised authorities, many major banks nowadays are also interested in using it to collaboratively settle transactions. \footnote{\url{https://www.reuters.com/article/us-blockchain-banks/six-big-banks-join-blockchain-digital-cash-settlement-project-idUSKCN1BB0UA}}

\textbf{iii) Blockchain Cost Models:} Croman et al. \cite{croman2016scaling} did a reality check of Bitcoin and analyzed the cost to confirm transactions. Rimba et al. \cite{rimba2017comparing} later proposed cost models using parameters such as \emph{gas} and \emph{gasPrice} to compare Ethereum with cloud services for business process execution. We stick to findings in \cite{croman2016scaling} to obtain deployment and operation costs as the gas-related parameters can be controlled by the bank.

\section{Discussion}
\label{sec:discussion}
By now we have demonstrated the design of a hybrid, blockchain-based payment scheme in an intermittently-connected environment. Although the ultimate control belongs to the bank, the reliable operation of the system is achieved via decentralization. Our system benefits all participants. Villagers are genuinely motivated to join as they can make low-fee, cash-less transactions efficiently. Miners are incentivised to work and receive rewards. By managing users' digital and fiat accounts, the bank can make a profit in a traditional way. While with a distributed, self-regulated blockchain for transaction processing, the bank can avoid single points of failure, and enjoys a lower deployment cost compared to building physical infrastructures and hiring experienced staff. However, some aspects of the system design can be further improved. 


\subsection{Blockchain Inefficiency}
PoW is a heavy consensus algorithm, and it becomes more inefficient in public cryptocurrencies as players compete against each other to earn profits. However, it does not appear to be a problem in our implementation with low-power mining nodes. One of the most promising replacements is the Proof-of-Stake (PoS) proposed by King et al. \cite{king2012ppcoin}. The idea is to assign mining rights based on the ``coin-age", which equals the currency amount times its holding period. More recent proposals include Bitcoin-NG \cite{eyal2016bitcoin} that utilizes ``leader selection" and ``micro-blocks", LocalCoin \cite{chatzopoulos2017localcoin} which operates on mobile devices without the Internet, and the Red Belly Blockchain \cite{crain2017leader} which incorporates a fast, leader-free Byzantine consensus. However, as PoS and other protocols also have vulnerabilities \cite{houy2014will}, we expect PoW to continue being the main consensus algorithm in blockchain-type systems in the near future.

\subsection{Security Considerations}

\subsubsection{Temporary Inconsistency (not a real threat)}
Even though \emph{blockchain forks} or \emph{stale blocks} are an important indicator of inconsistency \cite{decker2013information}, if everyone behaves honestly, eventually all conflicts can be resolved. Hence, stale blocks themselves are not a direct threat to network security. However, they increase the chance of \emph{double-spend} attacks that are usually caused by disagreements within the network \cite{gervais2016security}.

\textbf{\emph{Countermeasure:}} As suggested by Karame et al. \cite{karame2012two}, one way to mitigate \emph{double-spend} attacks is to apply a ``listening period" of a few seconds on the recepient side. This can be integrated into the mobile wallet design. 

\subsubsection{Network Partitioning (a real threat)}
Wust et al. \cite{wust2016ethereum} managed to perform an eclipse attack on a private ethereum blockchain and Natoli et al. \cite{natoli2016balance} leveraged the network partitioning to perform the balanced attack. 

\textbf{\emph{Countermeasure:}} In our private blockchain setting, the permitted miners may turn malicious at a later time. To enhance security, we can enable the bank to deploy a smart contract that tracks node connections and detects network partitioning. We here present a simple method described as follows. We start off with a randomly selected miner $m_1$ belonging to the set $N=\left\{ m_1, m_2, m_3, \cdots, m_{l_m} \right\}$. Miners that are connected directly to $m_1$ belong to the set $P_1 = \left\{ m_2, m_3, m_4, \cdots, m_k \right\}$. For each element in $P_1$, we obtain similar sets of $P_2$, $P_3$, etc. This process is repeated until we find a set $P_x \in P_1\cup P_2 \cup \cdots \cup P_{x-1}$. If there are still remaining elements, i.e. $P_x \subset N$, a subgraph is detected. When the bank is connected, it accesses the results generated by this smart contract. The bank can then take action, e.g. to force nodes in the subgraph to restart and establish new connections. Note that this method does not prevent attacks or take effect in real-time.

\section{Conclusion} 
\label{sec:conclusion} 
In this paper, we propose a blockchain-based payment scheme in a remote region setting which has intermittent connectivity to the bank's central system. 
To do so, we have introduced a novel way of using smart contracts by making them responsible for the admission control of the system, account management, mining rewards, and token creation. 
We developed mathematical models to analyze system dynamics and the costs for setup and operations. We validated the proposed models on a private Ethereum testbed and demonstrated the practicality of system design using laptops and mobile devices. 
Through a comprehensive study of the local blockchain operation, we showed its stability under network disturbances. 
We then demonstrated the whole system operates well in resource-constrained, dynamic, intermittently connected environments. 
Future work will address the security issues further, but we are confident that the security threats will not be a major bottleneck in realizing our proposal. 

\bibliographystyle{IEEEtranS}
\bibliography{references}

\end{document}